\documentclass[aps,paper,twocolumn]{revtex4}
\usepackage{amsmath}
\usepackage{amsfonts}
\usepackage{color}
\usepackage{amssymb}
\usepackage{graphicx}
\usepackage{epstopdf}

\begin{document}


\title{Comparing the influence of
distinct kinds of temporal disorder in a low dimensional absorbing
transition model}
\author{C. M. D. Solano$^1$, M. M. de Oliveira$^2$ and C. E. Fiore$^1$}
\address{
$^1$ Instituto de F\'isica, Universidade de S\~ao Paulo, 
S\~ao Paulo-SP,  05314-970, Brazil, \\
$^2$Departamento de F\'{\i}sica e Matem\'atica,
CAP, Universidade Federal de S\~ao Jo\~ao del Rei,
Ouro Branco-MG, 36420-000 Brazil. 
}

\date{\today}

\begin{abstract}
Recently one has stated  that  temporal disorder constitutes
a relevant perturbation in absorbing phase transitions for all dimensions. 
However, its effect for systems  other than the standard contact process 
(CP),  its competition with other ingredients (e.g. particle diffusion)
 and other kinds of disorder (besides the standard types) are unknown. 
In order to shed some light in the above mentioned points,
we investigate a variant of the usual CP, namely triplet annihilation
model (TAM), in which the competition between 
triplet annihilation and single particle diffusion leads to an unusual phase diagram behavior,
with reentrant shape and endless activity for sufficient large diffusion
rates. Two kinds of time-dependent disorder have been considered.
In the former, it is introduced in the creation-annihilation parameters (as commonly
considered in recent studies), whereas in the latter the diffusion rate $D$ 
(so far unexplored) is allowed to be time dependent. 
In all cases, the disorder follows an uniform distribution
with fixed mean and width $\sigma$. Two values of $\sigma$ have been considered, 
in order to exemplify the regime of ``weaker" and ``stronger" temporal
disorder strengths. Our results show that in the former approach,  the disorder suppresses the reentrant
phase diagram with a critical behavior deviating from the directed percolation universality class  (DP) in the regime of low diffusion rates, 
while they strongly suggest that 
the DP class is  recovered for larger hoping rates. 
An opposite scenario is found in the latter disorder approach, 
with a substantial increase of reentrant shape and the maximum diffusion, 
in which the reentrant shape also displays a critical behavior
consistent to the DP universality class 
(in similarity with the pure model). 
In order to compare with
very recent claims, results by taking a bimodal distribution
and critical behavior in the limit of strong disorder are presented. 
Also, results derived from the mean field theory (MFT) are performed, presenting
partial  agreement with 
numerical ones. Lastly, comparison with the diffusive
disordered CP has been undertaken.

\end{abstract}

\maketitle
\section{Introduction}

Although typical nonequilibrium phase transitions transitions into an absorbing phase belong generically to the very well established directed percolation
(DP) universality class \cite{marr99,odor07,henkel,hinrichsen,odor04}, the inclusion of impurities or
defects drastically affects its critical behavior 
\cite{noest,mordic,hooy,vojta,barg,vojta2,marc}. Such disorder approach is an unavoidable ingredient 
of real systems and has been argued as one of the causes for the rarity of DP behavior in
experiments \cite{exp}. For these reasons, the study of disordered systems has deserved considerable attention in the last years.

Impurities and defects are different kinds of spatial disorder
that leads to the existence of rare
regions in the absorbing phase, characterized  by
large local activity and slow decay (algebraic, instead of exponential)
toward the extinction with non-universal exponents \cite{vojta,barg}.
It is typically introduced under two distinct ways, 
through random spatial variation of the control parameter
\cite{vojta}  or random deletion of sites or bonds 
\cite{mordic,vojta2,marc}.  

Even though less studied than the spatial disorder,  
the temporal disorder has also attracted interest 
\cite{jensen96,kamenev,munoz2011,neto,oliv-fiore16,neto2}. In contrast to the previous
case, the control parameter is allowed to be time dependent, 
resulting in  temporarily active (ordered)
and absorbing (disordered) phases,
whose effects are more relevant  at the emergence of the phase transition. 
 Heuristically the importance of temporal disorder can be set through the generalization of 
Harris criterion (valid for spatial disorder) \cite{harris}, proposed by Kinzel \cite{kinzel}. According to it, 
the temporal disorder is a relevant perturbation whenever the clean  
temporal correlation exponent (without disorder)
$\nu_{||}$ satisfies the condition $\nu_{||}<2$. Since 
for the DP universality class $\nu_{||}$ reads $\nu_{||}=1.733847(6),
1.2950(60)$ and $1.110(10)$
in one, two and three dimensions respectively,  the   
 temporal disorder is expected destabilizing the 
DP absorbing phase transition.  Recent  
results for the contact process (CP), the prototypical model in the DP class, 
have confirmed this.  The temporal disorder  also reveals a region in the active phase, named
\emph{temporal Griffiths phase}, in which the mean lifetime 
exhibits a power-law behavior (instead of  exponential growth) 
\cite{munoz2011,neto,neto2}.
Despite this,  results are  restricted mainly  for 
the contact process (CP) and with partial conflicting conclusions. 
The first set of numerical results, from Refs. \cite{munoz2011,jensen96}, predict non-universal
critical behavior with algebraic behavior $t^{-\alpha}$ (being
$\alpha$ the critical exponent of a given quantity)
and dependent on the disorder strength.
On the other hand, based on a 
strong noise normalization group (RG) analysis and simulations, 
a behavior of type $t^{\theta}[\ln t]^{-\alpha}$ 
with universal critical exponents has been argued for the CP, in very recent works 
by Vojta et al. \cite{neto} 
and Barghathi et al. \cite{neto2}. 

Additionally, the effects of temporal disorder in the presence of
other ingredients, 
such as particle diffusion, and interactions have not been considered yet.
In particular, several works have 
shown that the diffusion constitute a relevant 
perturbation, affecting drastically
the critical behavior \cite{henkel,pcpd} or even
 leading to distinct scenarios for discontinuous
phase transitions \cite{fiore14,fiore15,scp2}.  

In order to cover the above points, in this work we
investigate the effects of temporal
disorder in a variant of the usual CP,  namely the triplet annihilation
model (TAM) \cite{dic89,fiore05}. It is a stochastic lattice model that was
 proposed a long time ago and it has been studied
 over distinct numerical approaches \cite{dic89,fiore05}.  Each lattice
 site in the TAM model
 is vacated or occupied by a particle. The creation events are similar
 to the CP, where 
single particles reproduce autocatalitically at rate $1$. 
But alternatively,  particle annihilation will occur
(with rate $\alpha$) only
if there are \emph{three} adjacent particles (the triplet is annihilated
and the three respective sites become empty). 
Although its critical behavior belongs
to the DP universality class (posing not any surprise), the inclusion 
of particle diffusion leads to an unusual phase diagram,
including reentrant shape and indefinite activity (i.e.
absence of phase transition) for sufficient high
diffusion rates from a maximum diffusion $D_{\rm max}^*$.
In the present work, we aim at covering three
fundamental points: (i) what are the effects of 
temporal disorder in distinct examples of 
absorbing phase transitions (besides
that describes the usual CP) as well as the influence of distinct
ingredients? (ii) what is the result of the distinct sorts
of temporal disorder? (iii) How do they compare with the CP?
These issues will be answered by considering
two kinds of temporal disorder, according
to uniform and bimodal distributions, respectively. 
Results show that the disorder considered
in the reactive part (creation-annihilation
parameters) leads to the  suppression (maintenance) 
of reentrant
phase diagram for the uniform (bimodal) cases 
and  critical behavior deviating from the DP in the regime
of low diffusion rates. On the other hand, irrespective
the disorder distribution 
numerical, our results (strongly) suggest that  the DP universality
 class  is recovered for larger values of the diffusion rates $D$.
In both cases, the maximum diffusion $D_{\rm max}^*$ is mildly affected 
by the disorder. An
 opposite scenario is resulted
when the disorder is included in the diffusion, with a substantial
increase of reentrant shape and $D_{\rm max}^*$.
A detailed analysis of the critical behavior in the regime 
of temporal disorder will be presented.

This paper is organized as follows: In Sec. II, we describe the model
and methods. In Sec. III we show the numerical results for the TAM,
whereas the comparison with the  disordered CP has been undertaken in Sec. IV.
Conclusions are presented in Sec. V.

\section{Model and methods}

In the TAM model, the creation of particles is similar to the usual CP 
(one  particle is required for creating
a new species in an empty neighboring site), but instead
only three adjacent particles can be annihilated (and the three sites are left vacant). 
Schematically, it  is represented by the chemical reactions $0+A\rightarrow^{1/2} A+A$
and $A+A+A\rightarrow^{\alpha} 0+0+0$, being  $1/2$ and $\alpha$
the creation and annihilation strength rates, respectively. Additionally,
with rate $D$ one species hops to one of its nearest-neighboring sites (chosen at random), provided
it is empty ($0+A\rightarrow^{D}A+0$). 

As mentioned previously, in the pure version (without disorder), the  
competition between diffusion and triplet annihilation
 changes dramatically the phase diagram,  including a reentrant portrait and no phase transition up to a limit diffusion rate $D_{\rm max}^{*}$ 
(in the one-dimensional case, $D_{\rm max}^{*} \sim 0.587$). For $D<D_{\rm max}^*$,
an active state is also possible for very large annihilation
rates. In Fig. \ref{fig5} (continuous lines), the phase
diagram for the pure version  is shown in terms
of the variable ${\sqrt \lambda_c}$, being  $\lambda_c=1/\alpha_c$.

The temporal disorder is introduced in form
so that the time changes by  $\Delta t=t^*$ (being $t^*$ 
the computational time) 
a given control parameter $y$ 
is updated to a new value. 
In the uniform case, the new
 $y$ is
extracted from an uniform 
distribution with mean $y_0$ and width $\sigma$,
implying that $y(t)=y_0+(2\xi-1)\sigma$, where 
$\xi$ is random number $\in [0,1]$. In the former approach, the 
creation-annihilation $y(t) \rightarrow p(t)$ changes with 
$p_0=1/(1+\alpha_0)$ (with fixed $D$),   whereas in the latter approach the diffusion 
is time dependent $y(t) \rightarrow D(t)$ (with fixed $p=1/(1+\alpha)$). 
Thus in the former (latter) cases,
$p(t)$ ($D(t)$) are restricted between 
$p_0-\sigma$ and $p_0+\sigma$ ($D_0-\sigma$ and $D_0+\sigma$).
Since $0\le p(t)\le 1$, the reactive
part  is composed only of  annihilation and creation
whenever $p(t)<0$ and $p(t)>1$, respectively.   
For the disorder included in the diffusion, the actual simulation
is composed of only creation-annihilation (diffusion) subprocesses, whenever
$D(t)$ has a negative (larger than 1) value.   
From now on we will call
the former (latter) case the $p$-disordered ($D$-disordered) TAM model.
In the bimodal case (studied only for the $p-$disordered TAM), 
as the time changes $\Delta t=t^*$
the  annihilation rate is chosen from two values,  
$\alpha$ and $\alpha_h>>\alpha$, with 
probabilities $1-{\tilde p}$ and ${\tilde p}$, respectively 
and $0<{\tilde p}<1$.

Numerical analysis will be carried out by performing
spreading experiments and the time decay of the system
density. In the first case,  one starts from an initial configuration
close to the absorbing state, except for three
adjacent particles placed in the center. 
Typical  quantities in this analysis are the survival probability $P_s(t)$,
the total number of particles $N(t)$ and the mean cluster size
$R^2(t)$. For usual absorbing phase transitions, at
the critical point they behave as
\begin{equation}
P_{s}(t) \sim t^{-\delta}, \quad N(t)\sim t^\eta \quad {\rm and }
\quad R^2(t) \sim t^{2/z},
\label{eq1}
\end{equation}
where $\delta,\eta$ and $z$ are their associated critical exponents. 
For the DP universality class they read $\delta=0.159464(6)$, 
$\eta=0.313686(8)$ and $z=1.580745(10)$ \cite{henkel}.

Second, the time evolution of 
the system density $\rho(t)$ starting from
a fully occupied lattice will be considered. At the critical point,  
$\rho(t)$ exhibits an power-law behavior 
\begin{equation}
\rho(t)\sim t^{-\theta},
\label{eq11}
\end{equation}
being $\theta$ its associated critical exponent.
By restricting the calculation only over survival 
runs, $\rho_{s}(t)$ also decays as
Eq. (\ref{eq11}) for the initial time but
 for larger $t$ it converges to a definite value $\rho_s(L)$. Finite-size scaling analysis
 shows that $\rho_s(L)$ depends on  $L$ according to 
 the scaling behavior $\rho_s(L) \sim L^{-\beta/\nu_{\perp}}$,
where $\beta$ and $\nu_{\perp}$ are the critical exponents associated 
to the order-parameter and the spatial length correlation, respectively. 
Since $\theta=\beta/\nu_{||}$ 
and $z=\nu_{||}/\nu_{\perp}$, above analysis 
not only provide two additional critical exponents, but also it
give us an achievement of $z$, in order to certify  the scaling 
behavior at the criticality.
For the DP universality class,  the rapidity-reversal symmetry
states the exponents $\theta$ and $\delta$ 
should be equal. This can be understood
by considering the DP process in which
 the occupation probability of the directed
percolation process is identical in both time directions 
(forward and backward), implying that the time evolution
of $\rho(t)$ starting from a fully active lattice  
is equivalent to the behavior of $P_s(t)$ in the reversal time direction.
Also, the hyperscaling 
relation $2(\theta+\delta+\eta)=2d/z$ is satisfied, as indeed verified
in Ref. \cite{jensen96}. 
Assuming a scaling of type predicted by Eq. (\ref{eq1}),
the  critical point can be located in an undoubtedly way 
by taking the behavior of $N(t)$ versus $P_s(t)$, 
in which the  dependence on time does not explicitly appear.  
More specifically, by isolating the time in $P_s(t)$ and 
substituting in the expression for $N(t$), 
 both kind of scalings provide the
similar expression $N(t)\sim P_s(t)^{-\alpha}$  (where
$\alpha$ is the ratio between the exponents
ruling the time evolution of $N(t)$ and 
$P_s(t)$, respectively). 
Off the criticality, $N(t) \times P_s(t)$ deviates from the algebraic behavior.
Such analysis also has revealed useful in the context of
spatial disorder \cite{vojta}, since it avoids the use of arbitrary
times for achieving the critical exponents.
For DP universality class one has $\alpha=\eta/\delta=1.968...$, whereas in
other cases, the values of $\alpha$ are expected to be different.

Recently, based on strong noise
RG group, Vojta et al. \cite{neto} and Barghathi et al. \cite{neto2}
portended a new kind of scaling in the presence of sufficient strong
temporal disorder (infinite noise) for the CP. According to it, relevant quantities
behaves as
\begin{equation}
  P_{s}(t) \sim (\ln t)^{-\bar {\delta}},\\ N_s(t)\sim t^\theta (\ln t)^{-y_N}\\,
   R(t) \sim t^{1/\bar {z}}(\ln t)^{-y_R}.
\label{eq3}
\end{equation}
The exponents $\bar{\delta}$, $\bar{z}$ have exact values
 $\bar{\delta}=1$, $\bar{z}=1$, whereas
$y_N$ and $y_R$ are unknown. In addition,
they  
satisfy the hyperscaling relation $y_N=2\beta/\nu_{\perp}+dy_R$,
where $\beta=1/2$, $\nu_{\perp}=1/2$ and for the one-dimensional
case $d=1$. The time evolution of system density is also given by
$\rho \sim (\ln t)^{-\bar {\delta}}$ and thus its exponent
is similar to the $P_s(t)$, in accordance with the time reversal symmetry.
Although the occupation probabilities changes, 
their values  in the forward and backward directions 
are equal (since $p(t)=p^*$ for $t_i<t<t_j$).    
Since the critical behavior in the presence of temporal disorder
is still unclear, we will discuss (compare) both kind of scalings
for distinct points in the phase diagram.

\section{Results}

\subsection{Temporal disorder in the creation-annihilation parameter: $p$-disordered TAM}

\begin{figure}
\includegraphics[scale=0.36]{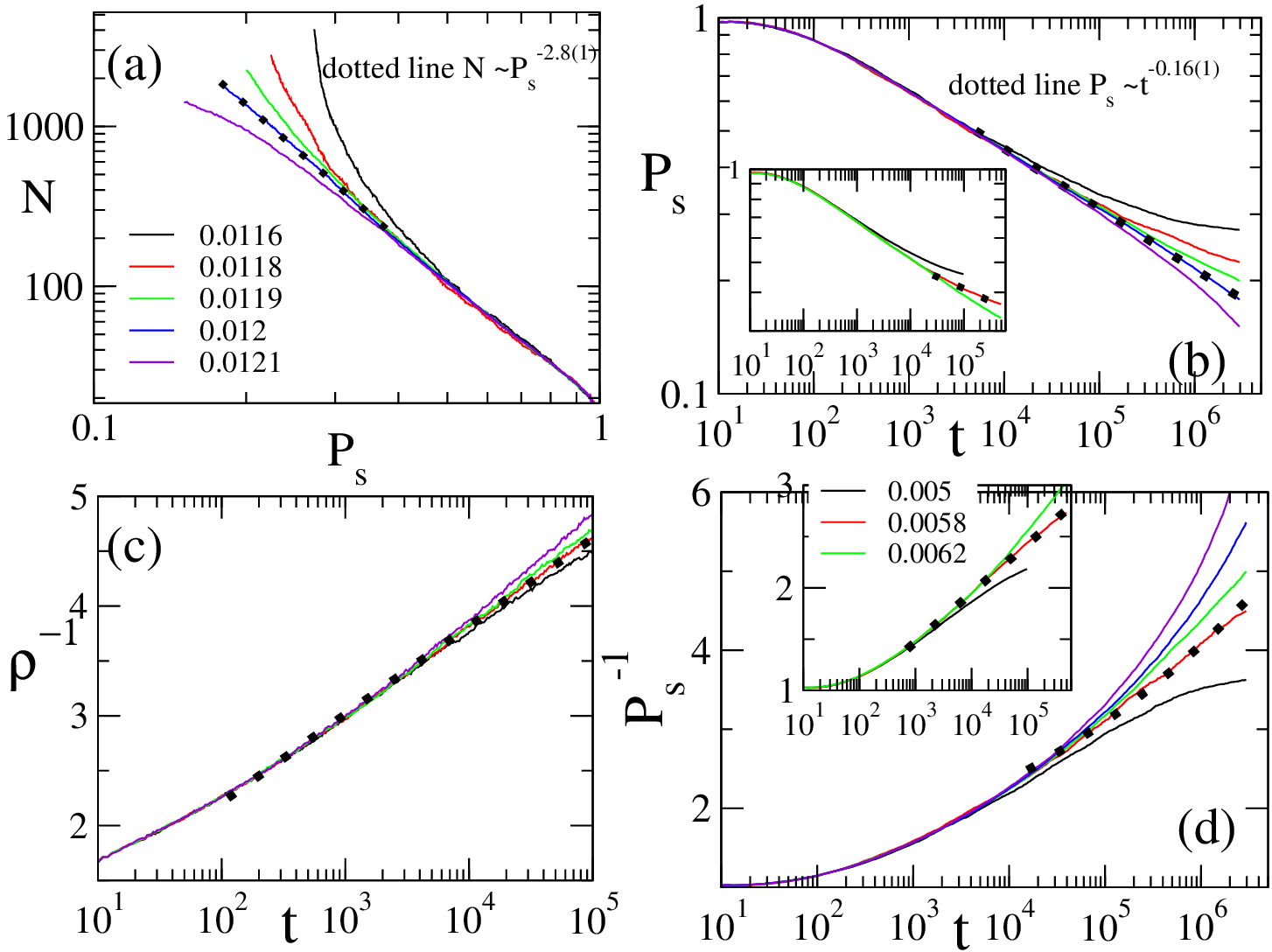}
\caption{({\bf Color online}): 
For the $p-$disordered TAM with
uniform distribution, results for $\sigma=0.5$, $D=0$
and distinct values of $\alpha_0$'s. In $(a)$ and $(b)$,
the log-log plot  $N$ vs $P_s$ and log-log plot of $P_s$ vs $t$
  for   $\Delta t=1$, respectively. 
  The dotted lines have slopes $\eta/\delta=2.8(1)$, $\delta=0.16(1)$,
  respectively.  
Panels $(c)$ and $(d)$ show, for the same $\alpha_0$'s,
  $\rho^{-1}$ and $P_{s}^{-1}$ vs $\ln t$.
The dotted lines have slopes consistent with Eq. (\ref{eq3}). Insets
show similar analysis, but for $\Delta t=2$. }
\label{fig1}
\end{figure}

\begin{figure}
\includegraphics[scale=0.36]{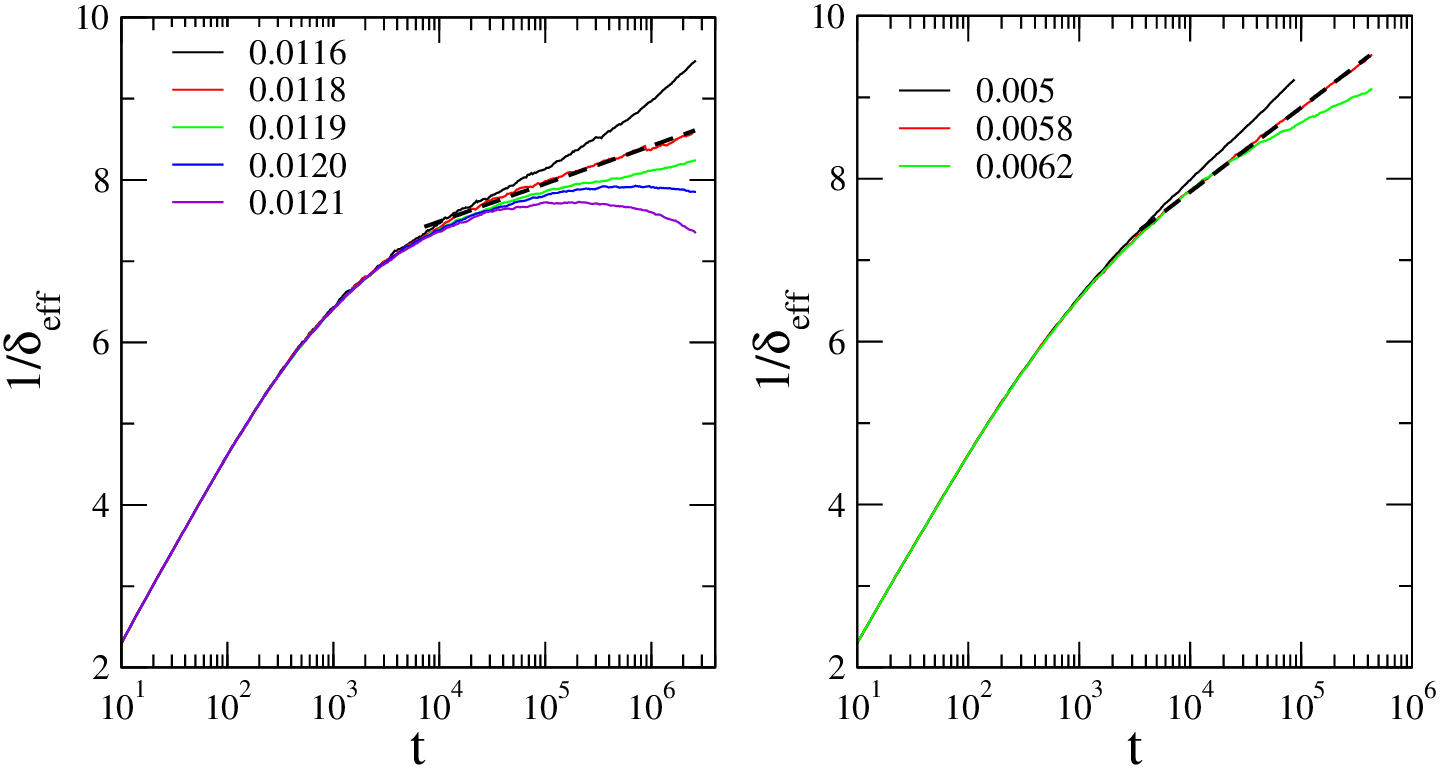}
\caption{({\bf Color online}):
For the $p-$disordered TAM with
uniform distribution,  the inverse of the effective exponent
$\delta_{\rm eff}$ for $\sigma=0.5$, $D=0$
and distinct $\alpha_0$'s. The left and right panels show results for  $\Delta t=1$  and 
$\Delta t=2$, respectively. Dotted lines correspond to the relation $\delta_{\rm eff} \sim (\ln t)^{-1}$.}
\label{eff}
\end{figure}

\begin{figure}
\includegraphics[scale=0.36]{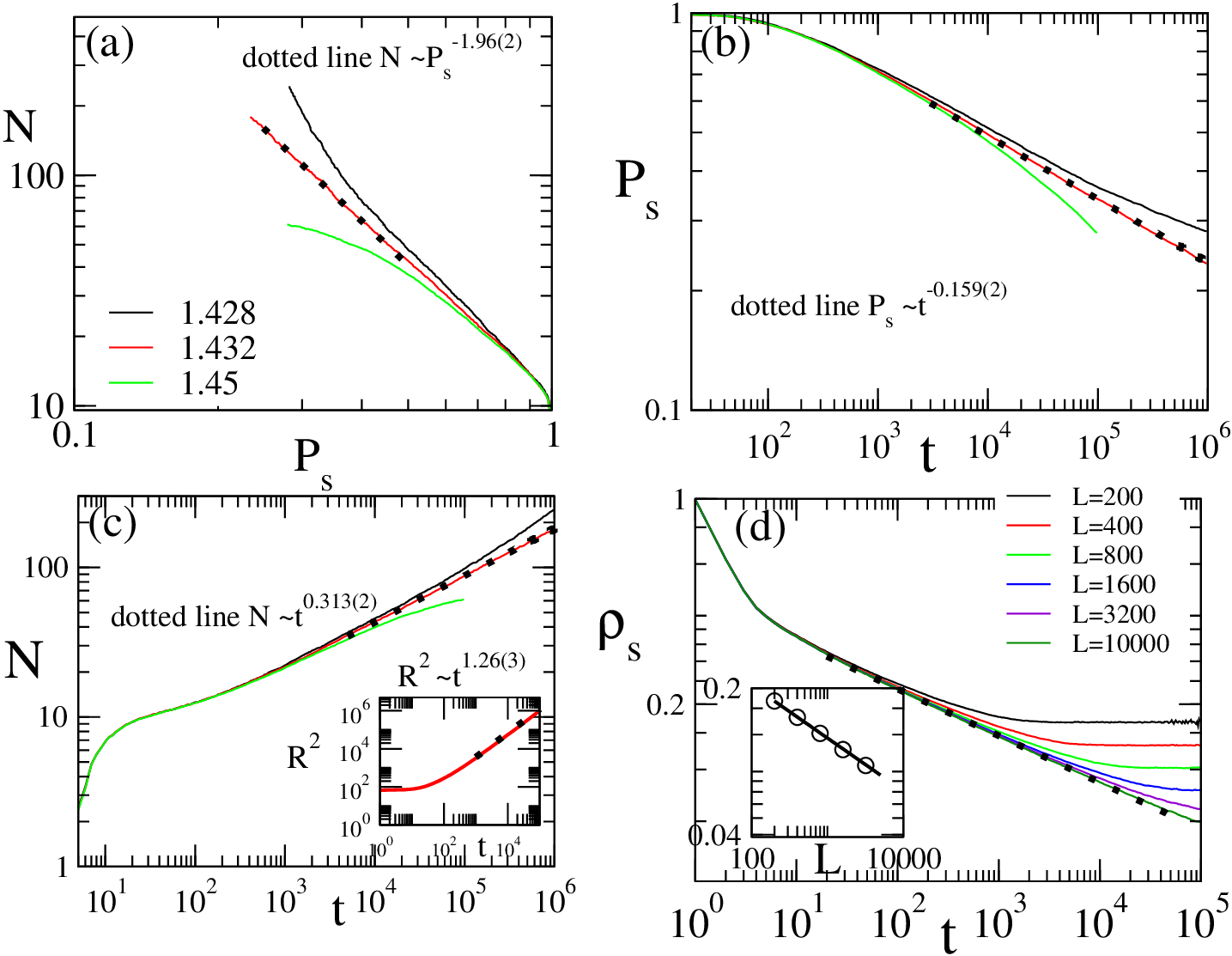}
\caption{({\bf Color online}): For the $p-$disordered TAM with
uniform distribution, results for $\sigma=0.2$ and $D=0.5$
and distinct $\alpha_0$'s. Panel
$(a)$ shows the log-log plot of $N$ vs $P_s$. In $(b)$ and $(c)$
the log-log plot of $P_s$ and $N$ as a function of the time,
respectively. Inset shows the $R^2$ at the critical point (red line).
In $(d)$ the log-log plot of
density over survival runs $\rho_s$ at $\alpha_{0c}\sim 1.432$. Inset:
log-log plot of $\rho_s$ vs $L$ and the straight
line has slope $\beta/\nu_{\perp}=0.252...$. 
The dotted lines have slopes  consistent with the
DP universality class. }
\label{fig2}
\end{figure}
In this section, we show the main results for the disorder
included in the creation probability ($p-$disordered TAM).
In all cases,  the averages will be evaluated by repeating
over $10^4-5\times10^4$ disorder configurations.
For a fixed rate  $D$,
the hoping, creation and annihilation process are chosen
$D$, $(1-D)p(t)$, $(1-D)(1-p(t))$, respectively.
In the first inspection, we analyze the critical
behavior for an uniform disorder
distribution for $D=0$, $\sigma=0.5$ and $\Delta t =1$,
as shown in Fig. \ref{fig1}.
As previous results for the usual CP 
\cite{munoz2011,neto2}, results  present a critical behavior
different from the DP universality class.
In particular, panels $(c)-(d)$
reveal that the behavior of $P_s(t)$ and $\rho(t)$ follow
the predictions from strong noise RG at $\alpha_{0c}=0.0118(1)$ 
for about three orders of magnitude in time,
 in conformity with Ref. \cite{neto2}. Conversely, 
panels $(a)-(b)$ show that the standard
critical behavior [Eqs. (\ref{eq1}) and (\ref{eq11})] is also verified
for about three orders of magnitude in time at $\alpha_{0c}=0.0120(1)$
(very close to the previous estimate). However, 
here the exponents $\theta$
and $\delta$ are different, suggesting that the time-reversal symmetry
is violated, although the hyperscaling relation 
$2(\theta+\delta+\eta)=2d/z$ has been verified within the error bars
(in the second decimal digit).  Extension
for other diffusion values is shown in Table
I, revealing a non-universal critical behavior in the regime of lower $D$'s.
Since the inclusion of temporal disorder is
not expected violating such symmetry (see previous discussion in Sec. II),  
we believe that Eq. (\ref{eq3}) predicts correctly the critical behavior
for larger $\sigma$'s (here exemplified 
 for $\sigma=0.5$) and low diffusion rates. With a view to 
strengthen the influence of temporal disorder, we repeat above analysis
 for $\Delta t=2$ (insets in Fig. \ref{fig1}) \footnote{Since the critical point decreases
greatly by increasing $\Delta t$, numerical results predicts 
the suppression of phase transition  
(the system is constrained in the absorbing phase for all values of
$\alpha_0$)  for $\sigma=0.5$ when $\Delta t \ge 3$.}.  
In fact, the crossover to Eq. (\ref{eq3})  is verified
for lower computational times ($t\sim 400$ instead of
$t=10^{4}$ for $\Delta t=1$), reinforcing that results
follow the  strong disorder RG \cite{neto,neto2}. 
Additional comparison between scalings is obtained by measuring the effective 
exponent $\delta_{\rm eff}$ from the relation $\delta_{\rm eff}=-d \ln P_s(t)/d \ln t$,
as shown in Fig. \ref{eff} for $\Delta t=1$ and $\Delta t=2$.
For  $\Delta t=1$,
although $\delta_{\rm eff}$ seems converging to a constant value at $0.0120$,
a decay of type $1/\ln t$  is verified for larger computational times at $0.0118$. Similar conclusion
is found for $\Delta t=2$ at $\alpha_{0c}=0.0058$. Thus, the evaluation of $\delta_{\rm eff}$
provide additional information about the validity of Eq. (\ref{eq3}).

 On other
hand, in the regime of weaker disorder (see e.g Table II for $\sigma=0.2$),
results do not exclude the critical behavior 
described by the usual scaling [Eqs. (\ref{eq1})-(\ref{eq11})], in conformity
with Refs. \cite{jensen96,munoz2011}.  We believe that 
much larger computational times are required for distinguishing both sort
of scalings, which are beyond our computational capacities. 
Despite this,  the shape of
phase diagrams are not affected by the difference
of scalings, since both criteria give  critical 
points very close to each other (differing in the fourth decimal digit).

\begin{figure}
\includegraphics[scale=0.36]{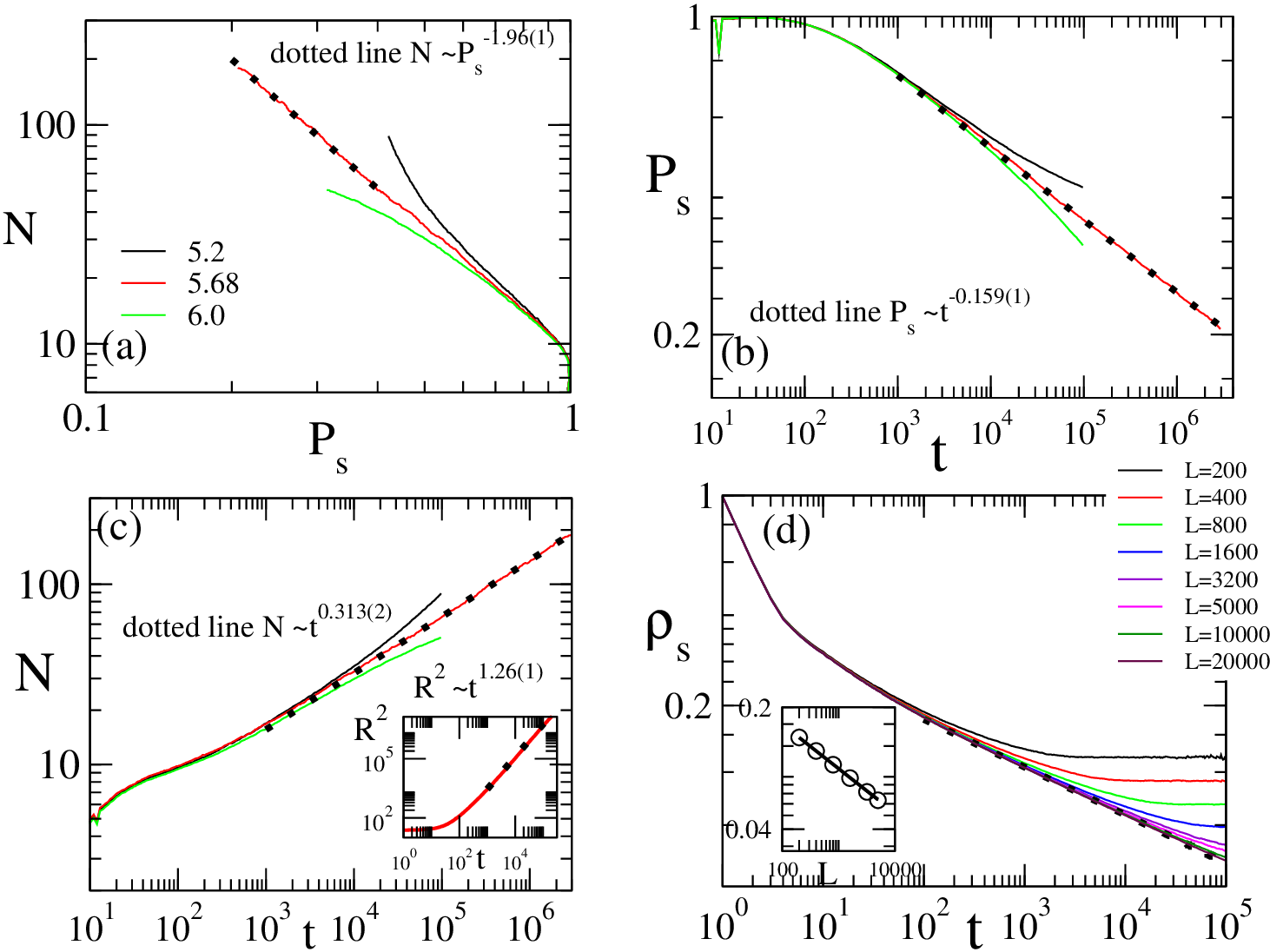}
\caption{({\bf Color online}): For the $p-$disordered TAM with
uniform distribution, results for $\sigma=0.5$ and $D=0.56$. 
In $(a)$, the log-log plot of $N$ vs $P_s$ for three distinct values of $\alpha_0$. In $(b)$ and $(c)$
the log-log plot of $P_s$ and $N$ as a function of the time,
respectively. Inset shows the $R^2$ at the critical point (red line).
 In $(d)$ the behavior of 
density over survival runs $\rho_s$ at the critical point 
$\alpha_{0c}=5.68$. Inset:
log-log plot of $\rho_s$ vs $L$ and the straight 
line has slope $\beta/\nu_{\perp}=0.252...$. 
The dotted lines have slopes consistent with the
DP universality class.}
\label{fig3}
\end{figure}

\begin{figure}[h]
\includegraphics[scale=0.35]{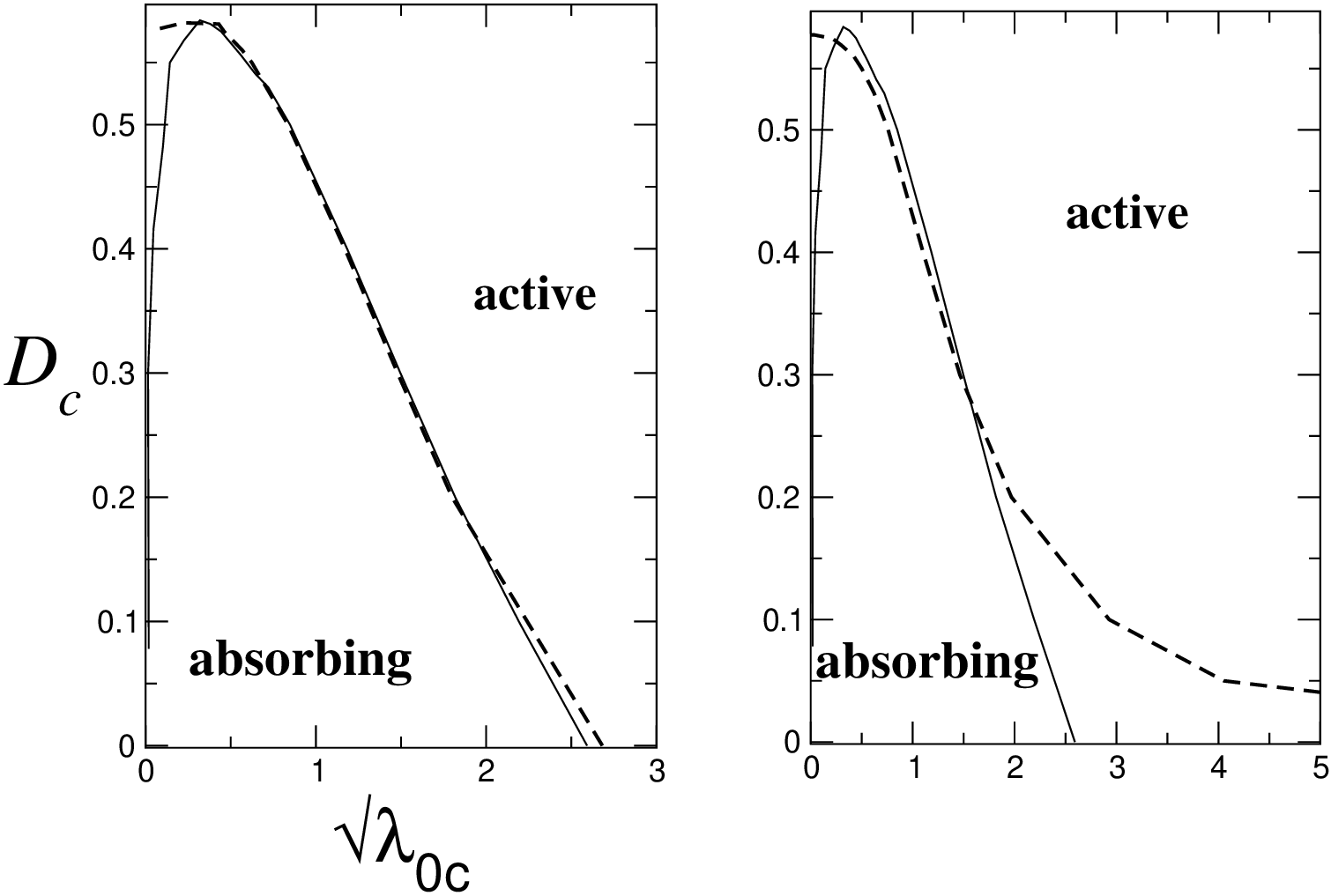}
\caption{Phase diagram for the 
$p-$disordered TAM (dashed lines) for $\sigma=0.2$ (left) and $\sigma=0.5$
 (right) and $\Delta t=1$. 
Continuous lines correspond to the pure model  
\cite{dic89,fiore05}.}
\label{fig5}
\end{figure}

An opposite scenario has been
verified in the regime of large diffusion rates.
In this case,  all critical exponents are consistent  
to the DP values, as shown in Tables I and II (larger $D$'s) and exemplified
in Fig. \ref{fig2}  and Fig. \ref{fig3} for  $D=0.5$ 
($\sigma=0.2$) and $0.56$ ($\sigma=0.5$),
respectively.
In both cases, different quantities follow power-law
behaviors consistent with their corresponding DP
exponents for at least three orders of magnitude in
time (see dotted lines). This provides the first evidence that
the  temporal disorder becomes irrelevant in the limit
of large diffusion rates. Fig. \ref{fig5}  shows the phase
diagram for both values of $\sigma$. In
order to compare with the results by Dickman (continuous lines) 
\cite{dic89}, they
are plotted in terms of ${\sqrt \lambda_{0c}}$, being 
$\lambda_{0c}=1/\alpha_{0c}$.
\begin{table}
\begin{ruledtabular}
\begin{tabular}{cccccccc}
$D_{c}$&$\alpha_{0c}$&$\delta$ & $ \theta$ &  $\eta$  & $2/z$&$\beta/\nu_{\perp}$   \\
\hline
0 &0.0120(1)&0.16(1)&0.11(1) &0.45(1)   &1.33(1)  &0.168(1) \\
0.2 &0.2579(1)&0.18(1) & 0.13(1) & 0.35(1)& 1.28(4)  &0.203(3)\\
0.53 &2.589(1) &0.161(2) & 0.157(2) & 0.317(3)& 1.26(3)&0.252(2)  \\
0.56 &5.68(2) &0.159(1) & 0.155(2) & 0.313(2)& 1.26(3)&0.253(2)  \\
DP&-&0.1594...&0.1594...&0.3136...&1.265...&0.252...
\end{tabular}
\end{ruledtabular}
\caption{For the $p-$disordered TAM with
uniform distribution, the set of critical exponents for $\sigma=0.5$ and 
distinct values of $D_c$ 
by taking an algebraic scaling from Eqs. (\ref{eq1}) and (\ref{eq11}).}
\label{table2}
\end{table}

\begin{table}
\begin{ruledtabular}
\begin{tabular}{cccccccc}
$D_c$&$\alpha_{0c}$&$\delta$ & $ \theta$ &  $\eta$  & $2/z$&$\beta/\nu_{\perp}$   \\
\hline
0 &0.1390(2) &0.15(1) &0.13(1) & 0.40(1) &1.32(3)  &0.199(1)  \\
0.2 & 0.3086(1)& 0.16(1) & 0.15(1) & 0.34(2) & 1.26(2)&0.233(2)  \\
0.5 &1.432(1) &0.159(2) &0.155(2) &0.313(2) & 1.26(4)& 0.253(2)  \\
0.55 &2.57(1) &0.159(1) &0.156(2) &0.313(2) & 1.26(2)& 0.253(2)  \\
0.575 &4.70(1) &0.159(2) &0.158(1) &0.314(2) & 1.26(2)& 0.253(2)  \\
DP&-&0.1594...&0.1594...&0.3136...&1.265...&0.252...
\end{tabular}
\end{ruledtabular}
\caption{For the $p-$disordered TAM with
uniform distribution, the set of critical exponents for $\sigma=0.2$ and distinct values of $D$
by taking an algebraic scaling from Eqs. (\ref{eq1}) and (\ref{eq11}).}
\label{table1}
\end{table}

The inclusion of the uniform distribution of 
disorder  induces the suppression of the reentrant shape, 
whose effects are more pronounced for $\sigma=0.5$. 
Although a very small reentrance is observed for
$\sigma=0.2$, it is absent for $\sigma=0.5$. 
 On the other hand, the maximum diffusion $D_{\rm max}^{*}$ separating
phase transition from endless activity is practically 
unaffected by the disorder, yielding at 
$D_{\rm max}^{*}=0.581(2)$ and $0.578(2)$- 
estimates very close to
the pure value $0.587$, once again validating the  irrelevance of
disorder  for sufficient large diffusion rates.

In the absorbing phase, we expect an asymptotic behavior of type 
$P_s(t) \sim e^{-\mu t}$, whose exponent $\mu$ 
depends on $\alpha_0$ and $D$.  
Starting from the critical point and increasing $\alpha_0$ (for a fixed $D$),
the exponent 
$\mu$ behaves slower for larger diffusion than for lower rates. In fact,
for $\sigma=0.5$ and $D=0$, the exponent   
$\mu$ are ranged in the interval $6\times10^{-5} \le \mu \le 7.5\times10^{-2}$ 
for $0.02\le \alpha_0 \le 36$,
whereas for $D=0.5$, 
we have that $4\times10^{-5} \le \mu \le 4\times10^{-4}$ (in the interval $2 \le \alpha_0 \le 36$).
Thus, the diffusion works against the absorbing phase
in the TAM, in which  $\mu \rightarrow 0$ 
for $D \rightarrow D_{\rm max}$, consistent to the emergence of
an endless active phase.

\begin{figure}
\includegraphics[scale=0.35]{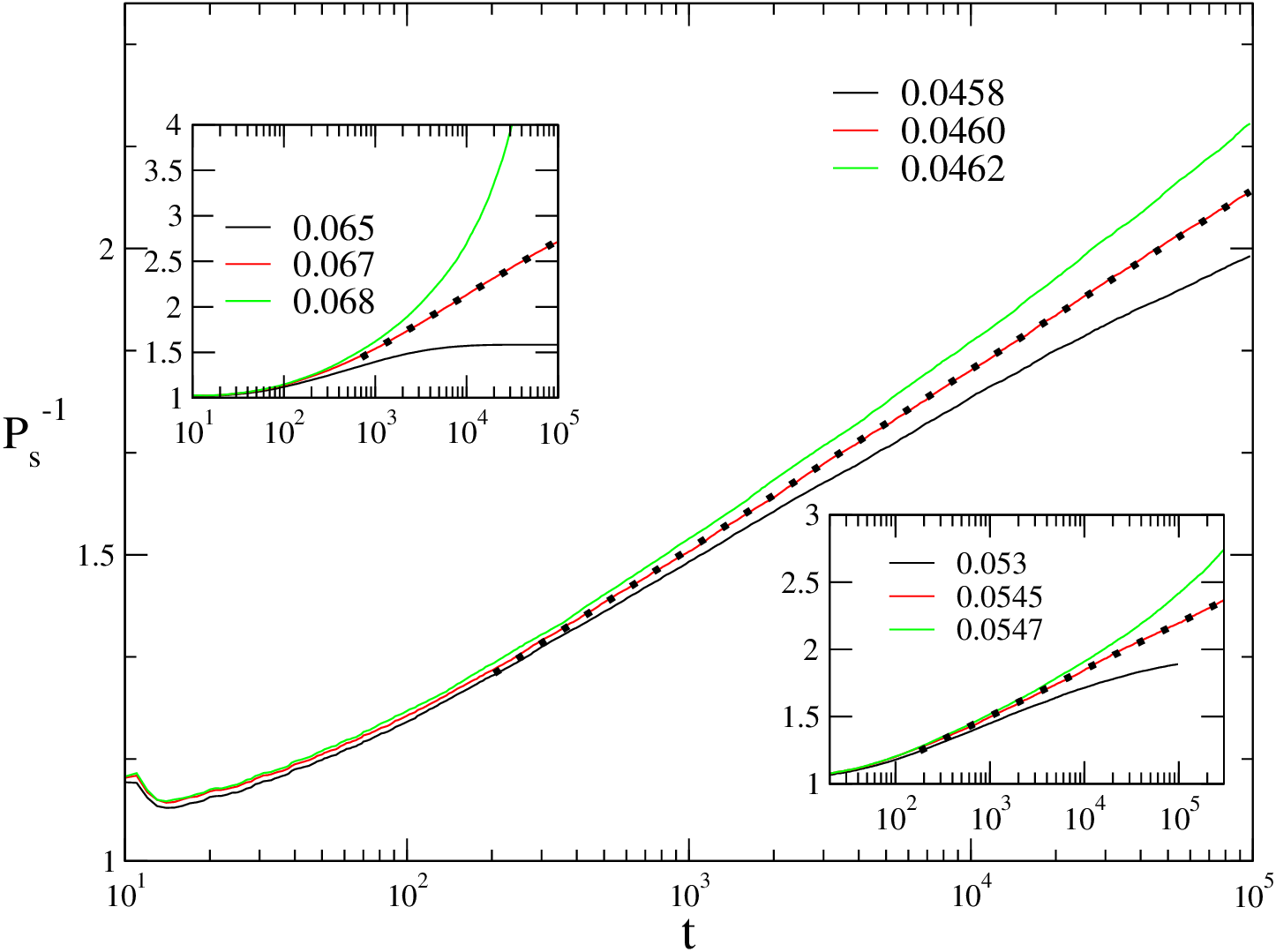}
\caption{({\bf Color online}): For the $p-$disordered TAM with
bimodal distribution,  $P_{s}^{-1}$ vs $\log t$ for 
 ${\tilde p}=0.2$, $D=0$, $\alpha_h=10\alpha$ and
$\Delta t=1$ (left), $\Delta t=6$ (right) and $\Delta t=10$ (center).
The dotted lines have slopes consistent $P_{s}^{-1} \sim \ln t$.}
\label{fig5-t}
\end{figure}

\begin{figure}[h]
\includegraphics[scale=0.35]{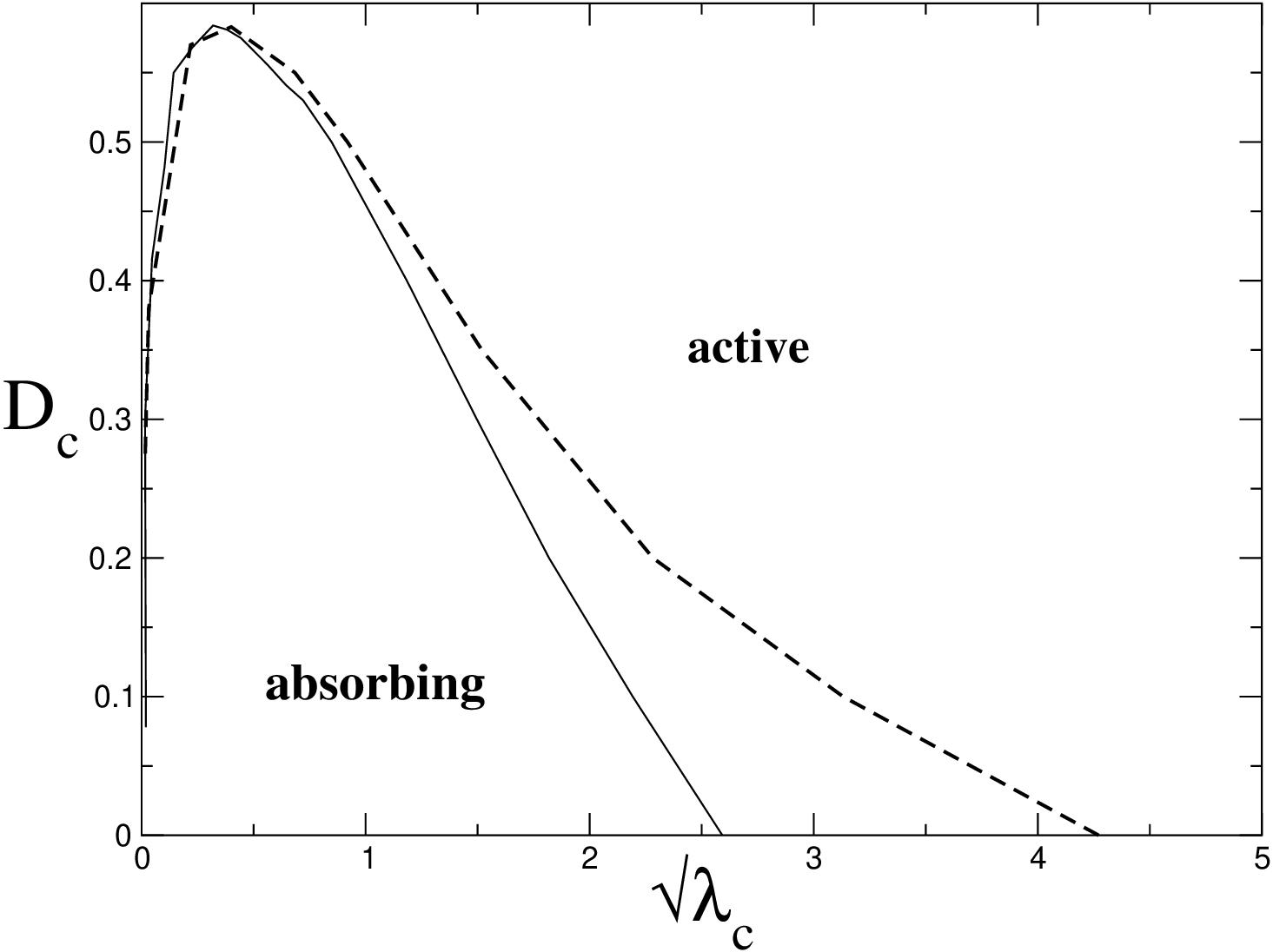}
\caption{The phase diagram for the 
  $p-$disordered TAM (dashed lines)
  for a bimodal distribution with $\Delta t=6$. 
Continuous lines correspond to the pure model
   \cite{dic89,fiore05}. }
\label{fig5-5}
\end{figure}

\begin{figure}
\includegraphics[scale=0.35]{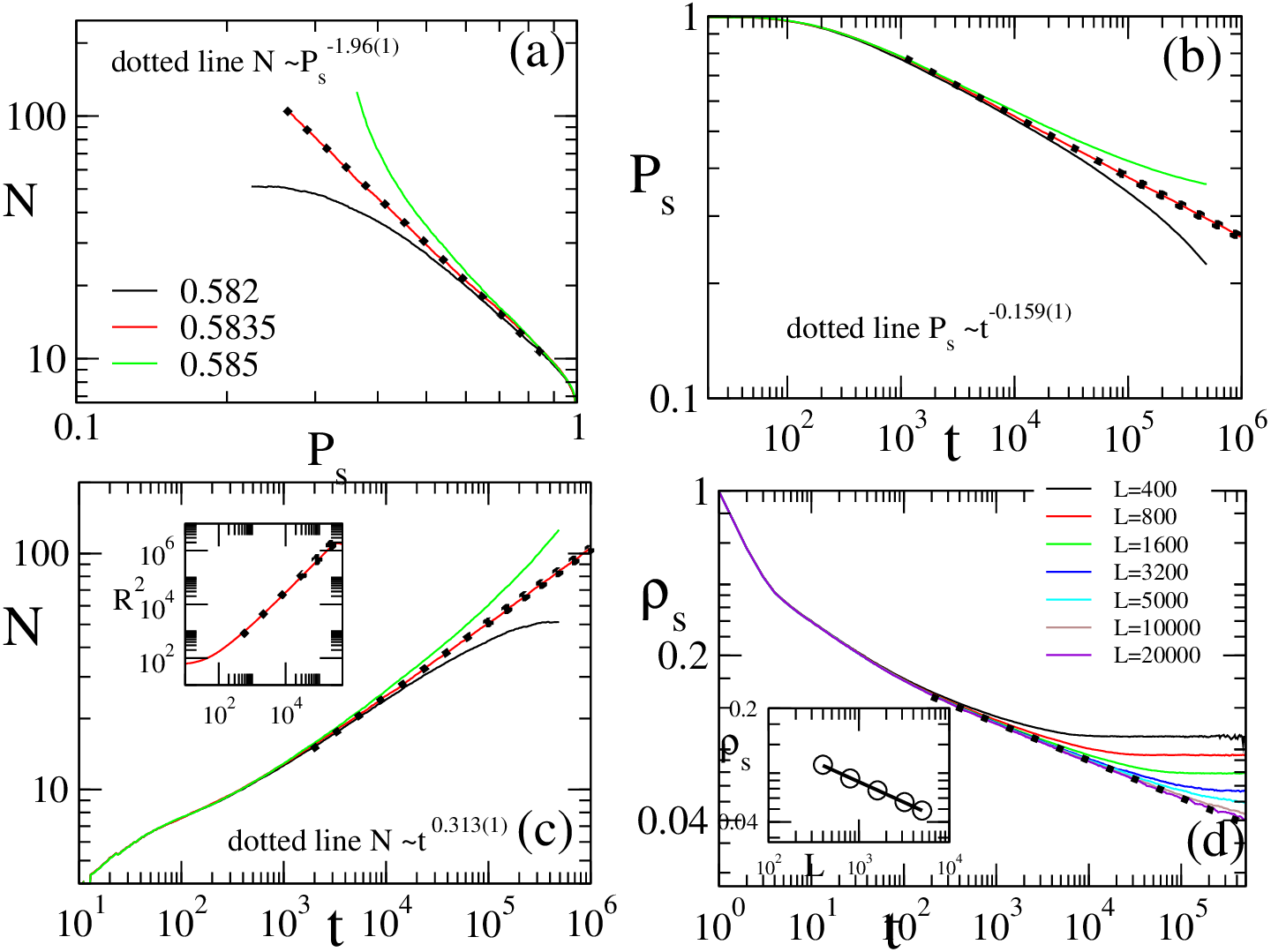}
\caption{({\bf Color online}):  For the $p-$disordered TAM with
bimodal distribution,  results for $\alpha=6.25$ and $\Delta t=6$. 
Panel $(a)$ shows the log-log of $N$ vs $P_s$ for three distinct values of $D$. 
In $(b)$ and $(c)$
the log-log plot of $P_s$ and $N$ as a function of the time,
respectively. Inset shows the $R^2$ at the critical point (red line).
The dotted lines have slopes consistent with the DP universality
class. In $(d)$ the behavior of
density over survival runs $\rho_s$ at the critical point $D_{c}=0.5835$. Inset:
log-log plot of $\rho_s$ vs $L$ and the straight
line has slope $\beta/\nu_{\perp}=0.252...$}
\label{fig5-6}
\end{figure}

In order to draw a comparison with  Refs. \cite{neto,neto2} 
as well as inspecting the influence of the disorder distribution,  
we consider the $p-$disordered TAM for a bimodal distribution.
Firstly, we examine the influence of $\Delta t$'s, as shown in Fig. \ref{fig5-t}
for $D=0$, $\alpha_h=10\alpha$ and ${\tilde p=0.2}$. 
As for $\sigma=0.5$, by increasing $\Delta t$
the crossover for the strong noise scaling is approached
for lower $t$'s.
 The phase diagram is shown in Fig. \ref{fig5-5} for ${\tilde p}=0.2$, 
$\alpha_h=10\alpha$ and $\Delta t=6$.
  As in the uniform case, the critical behavior deviate from
  DP for small diffusion rates, following the scaling consistent
 to Eq. (\ref{eq3}). Also, for $D > 0.5$, results also propound a  critical
behavior consistent  to the DP universality class (see e.g. Fig. \ref{fig5-6}). 
Note that as $D$ increases the critical
point becomes very close to the pure case, conjecturing once again the
smallness of temporal disorder in the limit of strong hoping rates.
On the other hand, the distinct kind of disorder leads to
some differences in the topology of phase diagram. Unlike
 the uniform case, the disorder 
does not suppress the reentrance shape. Also, for
a  given diffusion, the disorder critical point $\alpha_c$ 
is always  lower ($\lambda_c$ is larger) 
than  its the pure value. 
This can be understood  in  
contrast to the bimodal distribution (in which the disorder makes
$p(t)$ be only lower than $p_0$),  
the uniform case provides the  probabilities 
$p(t)$ to be larger and lower the $p_0$. 
A last comment concerns that although  $\Delta t$ is a relevant
parameter for low diffusion (see e.g Fig. \ref{fig5-t}), this
is not the case for larger $D$'s. For example, for 
$\Delta t=1$ and $\alpha_c=6.25$, the critical point yields
at $D_c=0.5842(2)$, very close to $D_c=0.5835(5)$
for $\Delta t=6$, thus closely related to the irrelevance
of disorder for larger hoping rates.

 A last comment concerns that  although the results  
suggest the DP behavior being recovered for large diffusion  (and thus
violating the Kinzel's criterion), one can not discard
a crossover to the scaling behavior predicted by Eq. (\ref{eq3}). 

\subsection{Temporal disorder in the diffusion parameter: $D$-disordered TAM}

\begin{figure}
\includegraphics[scale=0.35]{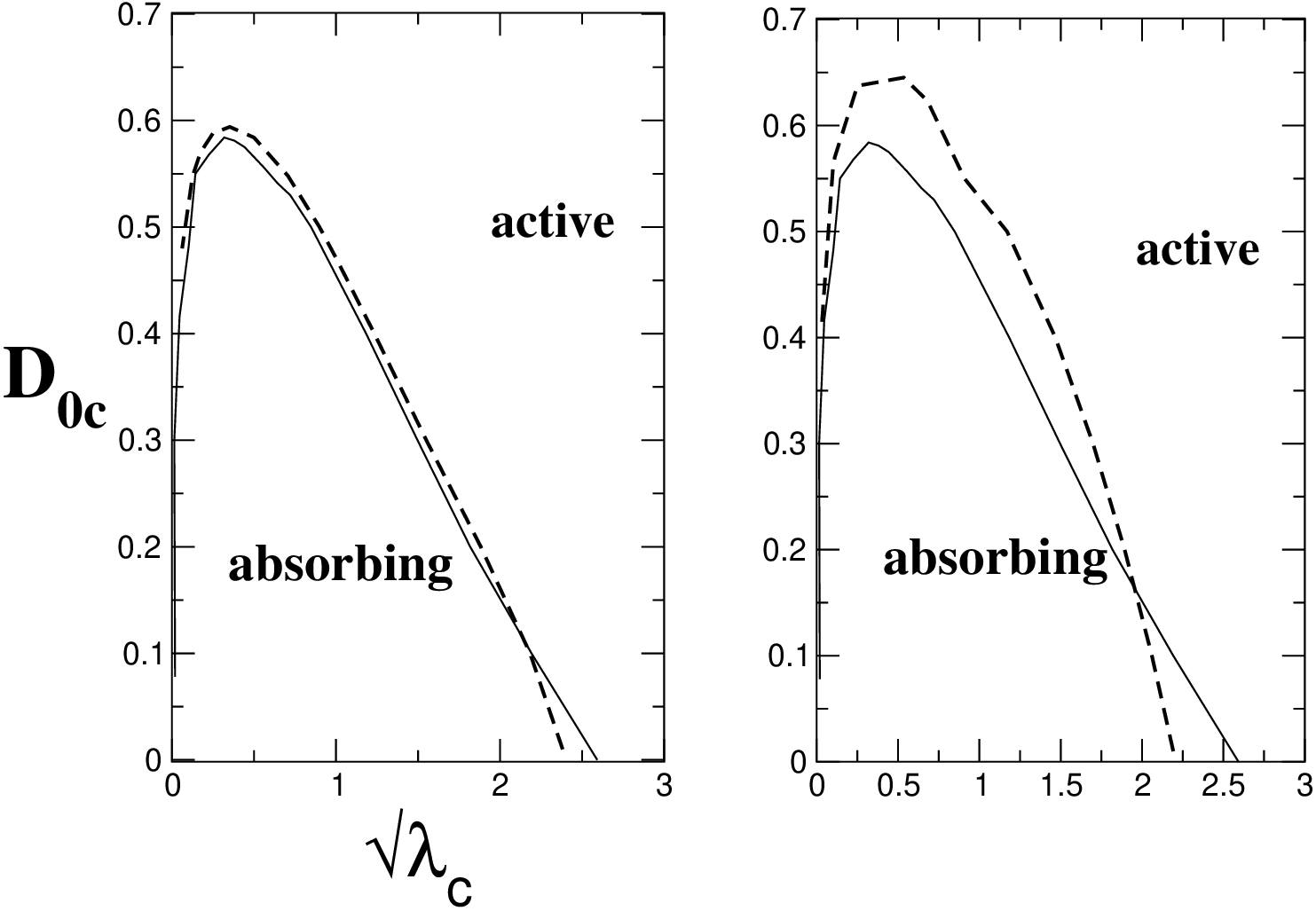}
\caption{Phase diagram for the 
$D-$disordered TAM  for $\sigma=0.2$ (left) and  $\sigma=0.5$ (right). 
Continuous lines correspond to the pure model \cite{dic89,fiore05}.}
\label{fig7}
\end{figure}

The temporal disorder is introduced in a similar manner, 
in which the diffusion rate is time dependent
with static creation-annihilation rates. Thus,
the diffusion, creation and annihilation processes are chosen 
 with probabilities  $D(t)$, $(1-D(t))p$, 
$(1- D(t))(1-p)$,  respectively being
$p=1/(1+\alpha)$ $D(t)=D_0+\sigma(2\xi-1)$
and $\xi$ also extracted from an random distribution $\in [0,1]$
whenever $\Delta t =t^*$.
In Fig. \ref{fig7}, we show the phase diagrams are shown  
for $\sigma=0.2$ and $0.5$ and $\Delta t=1$. Surprisingly, unlike the
previous (uniform) cases, the $D-$disordered version not only
maintains the  reentrant part, but also enlarges
the  absorbing phase and the maximum diffusion $D_{\rm 0max}^{*}$,
whose  effects are   more pronounced for $\sigma=0.5$.
In particular, we found the $D_{\rm 0max}^{*}$ reads $0.595(5)$ and $0.65(1)$,
for $\sigma=0.2$ and $0.5$, respectively.

\begin{figure}
\includegraphics[scale=0.35]{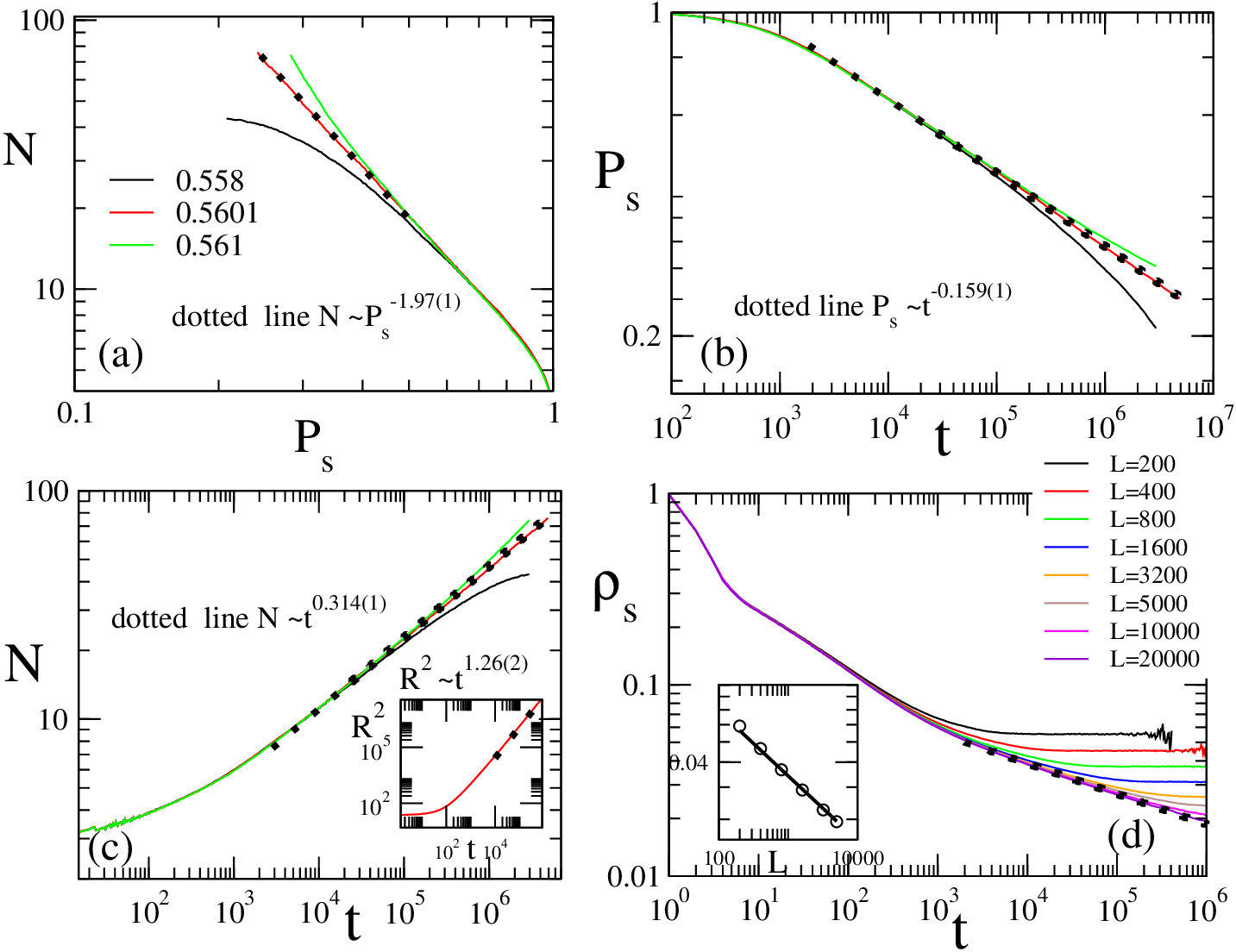}
\caption{({\bf Color online}):  
For the $D-$disordered TAM  
with uniform distribution, 
results for $\sigma=0.5$, $\alpha=100$, $\Delta t=1$
and distinct $D_0$'s. In $(a)$, the log-log plot of 
$N$ vs $P_s$ and in panels $(b)$ and $(c)$, 
the log-log plot of $P_s$ and $N$ as a function of the time,
respectively. Inset shows the $R^2$ at the critical point (red line).
In $(d)$ the behavior of
density over survival runs $\rho_s$ at the critical point  $D_{0c}=0.5601$. Inset:
log-log plot of $\rho_s$ vs $L$ and the straight
line has slope $\beta/\nu_{\perp}=0.253(2)$. The dotted lines have slopes consistent with the
DP universality class for more than three decimal levels. }
\label{fig6-10}
\end{figure}

\begin{figure}
\includegraphics[scale=0.35]{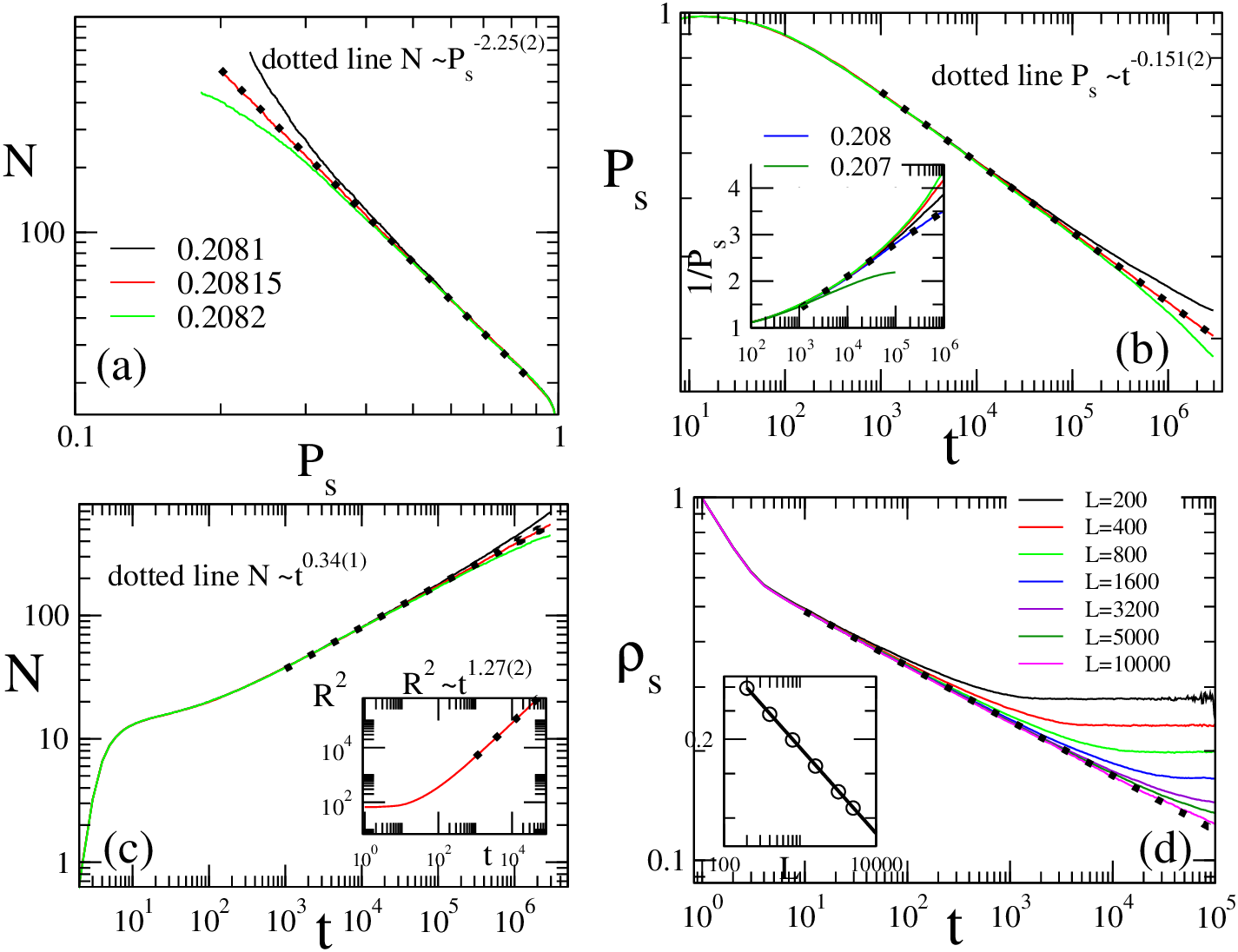}
\caption{({\bf Color online}): 
For the $D-$disordered TAM  
with uniform distribution,results for $\sigma=0.5$, $\Delta t=1$ and $D_0=0.01$.
Panel $(a)$ shows the log-log plot of $N$ vs $P_s$ for three distinct values of
$\alpha$. In $(b)$ the log-log plot of $P_s$ vs $t$.
For comparison, inset shows the monolog plot of 
$P_{s}^{-1}$ vs $t$. 
In $(c)$ the log-log plot of $N$ vs $t$ and
 inset shows the $R^2$ at the critical point (red line).
 Panel $(d)$ shows the log-log plot
of $\rho_s$ vs $t$. 
at $\alpha_c=0.20815$.
Inset: log-log plot of $\rho_s$ vs $L$ and
the straight line has slope $\beta/\nu_{\perp}=0.245(5)$. The dotted lines have slopes $\eta/\delta=2.25(2)$, $\delta=0.151(2)$,
$\eta=0.34(1)$, $2/z=1.27(2)$ for more than three decimal levels.}
\label{fig4}
\end{figure}

Table III exemplifies the critical exponents for 
distinct diffusion values and $\sigma=0.5$ by assuming
the usual scaling behavior. As
exemplified in Fig. \ref{fig6-10}, results suggest a
behavior consistent to the DP universality class
in the reentrant shape (for both $\sigma$'s). 
Outside this region, the critical
behavior clearly deviate from the DP. For $\Delta t=1$, the usual power-law 
behavior describes the results somewhat better than the strong noise RG
\cite{neto2}. This is exemplified in Fig. \ref{fig4} [panels $(b)$
and inset], in which results are well described by Eqs. (\ref{eq1})
and (\ref{eq11}) for more than three decimal levels. The exponents
$\delta$ and $\theta$ are also equivalent (within error bars). 
Similar results are obtained for distinct points of
the phase diagram (for both $\sigma$), although critical exponents
are rather few precise (uncertainties  
are in the second decimal digit). Thus,
we believe that scaling from Eq. (\ref{eq3}) (see inset in $(b)$) can not
be discarded.  Comparison with the critical behavior
for $\Delta t=6$ (not shown)  reveals that, in contrast to the $p-$disordered
case, there is no substantial crossover to the strong-noise RG theory
for lower times.
\begin{table}
\begin{ruledtabular}
\begin{tabular}{cccccccc}
$D_{0c}$&$\alpha_c$&$\delta$ & $ \theta$ &  $\eta$  & $2/z$&$\beta/\nu_{\perp}$  \\
\hline
0.01 &0.20815(5)&0.151(5) &0.151(5) & 0.34(1) &1.27(2)  &0.245(5) \\
0.6452(1) &3.5 &0.13(1) & 0.15(1) & 0.34(2)& 1.23(2)  &0.249(1)\\
0.6231(1)&2.15&0.13(1)&0.15(1) &0.35(1)&1.25(1)&0.241(2)&\\
0.6373(1)&16&0.158(2)&0.154(1)&0.314(2)&1.24(2)&0.256(2)\\
0.5601(1) &100 &0.159(1) & 0.151(2) & 0.314(1)& 1.22(3)&0.253(2)\\
0.411(1)&1000&0.160(1)&0.15(1)&0.312(1)&1.24(2)&0.25(1)\\
DP&-&0.1594...&0.1594...&0.3136...&1.265...&0.252...
\end{tabular}
\end{ruledtabular}
\caption{For the $D-$disordered TAM 
with uniform distribution, the set of critical exponents for $\sigma=0.5$ and 
distinct values of $D_{0c}$ by taking an algebraic scaling from Eqs. (\ref{eq1}) and (\ref{eq11}).}
\label{table2}
\end{table}

\subsection{Temporal Griffiths Phases (TGPs)}
\begin{figure}
\includegraphics[scale=0.35]{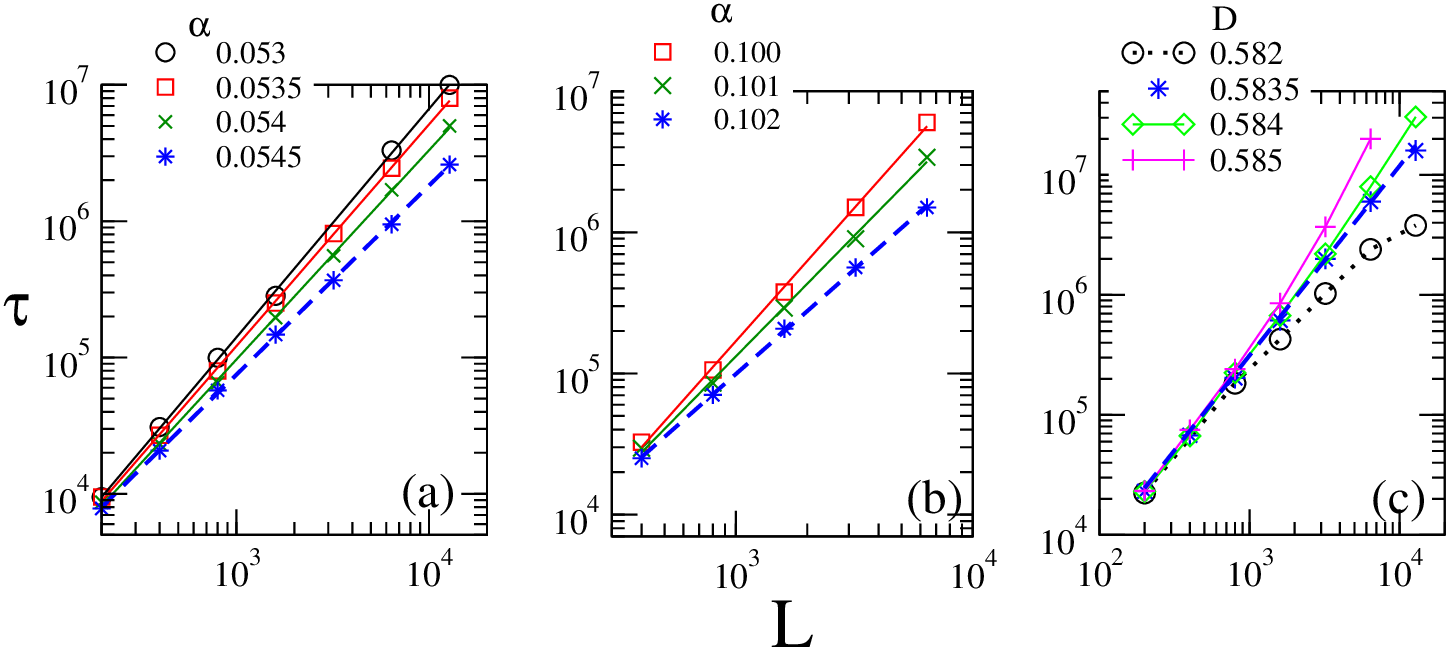}
\caption{({\bf Color online}): For the $p-$disordered TAM  
with bimodal distribution, the log-log plots of
 $\tau$ versus  $L$ for distinct $\alpha$'s and
$D$'s for panels $(a)-(b)$ and $(c)$, respectively. Panels 
$(a)$, $(b)$ and $(c)$ correspond to $D=0$, $0.1$  and $\alpha=6.25$,
respectively.  The dashed,  continuous and dotted lines correspond to the 
critical points, active and absorbing phases, respectively. Note that
panels $(a)-(b)$ show  a region in the active behavior,
in which $\tau$ grows algebraically with $L$, which is  not  presented in $(c)$.  }
\label{fig7t}
\end{figure}

Although Vasquez \cite{munoz2011} have found
the presence of TGPs in the CP only
for $d \ge 2$,  recent results by
Barghathi et al. \cite{neto2} have verified their occurrence 
in $d=1$ for larger $\Delta t$.
In this section, its presence 
for the $p-$disordered TAM has been investigated. 
For instance, we evaluate the mean life time $\tau$ 
by measuring the spent time to the system  
 reaching the absorbing phase (starting from
a fully occupied initial configuration). The $\tau$ is then obtained by repeating
such process over a sufficient number of identical initial configurations 
(here we used $10^3$).
Our analysis focus on the bimodal distribution for
$\Delta t=6$ and three distinct points of the phase diagram.
Fig. \ref{fig7t} shows the main results for $D=0$, $0.1$ and $\alpha=6.25$
with corresponding critical points reading $\alpha_c \sim 0.0545$, 
$0.102$ and $D_c \sim 0.5835$, respectively.

Note that for $D=0$ and $D=0.1$, there is a clear region
in the active phase (continuous lines) 
in which   $\tau$ exhibits  algebraic dependence
on $L$  displaying with non-universal exponents
(increasing  by lowering $\alpha$).
In particular, at the critical
points (dashed lines) the slopes read $1.3(1)$, $1.4(1)$ in 
rather few agreement with the theoretical value $1$. Thus,
as for the CP, TGPs are presented for the TAM. 
 Opposite  results are seen for the larger diffusion example 
(and $\alpha=6.25$).  No region within the active phase
with algebraic behavior is presented. Only  
at the criticality  $\tau$ behaves as $\tau \sim L^{1.59(1)}$, 
consistent  to the DP universality class exponent 
$z=\frac{\nu_{||}}{\nu_{\perp}}=1.5807...$ . Thus, 
the analysis of $\tau$ for finite systems suggest once again
the irrelevance of temporal disorder for larger diffusion values.
 
\subsection{Mean-field results}
\begin{figure}
\includegraphics[scale=0.3]{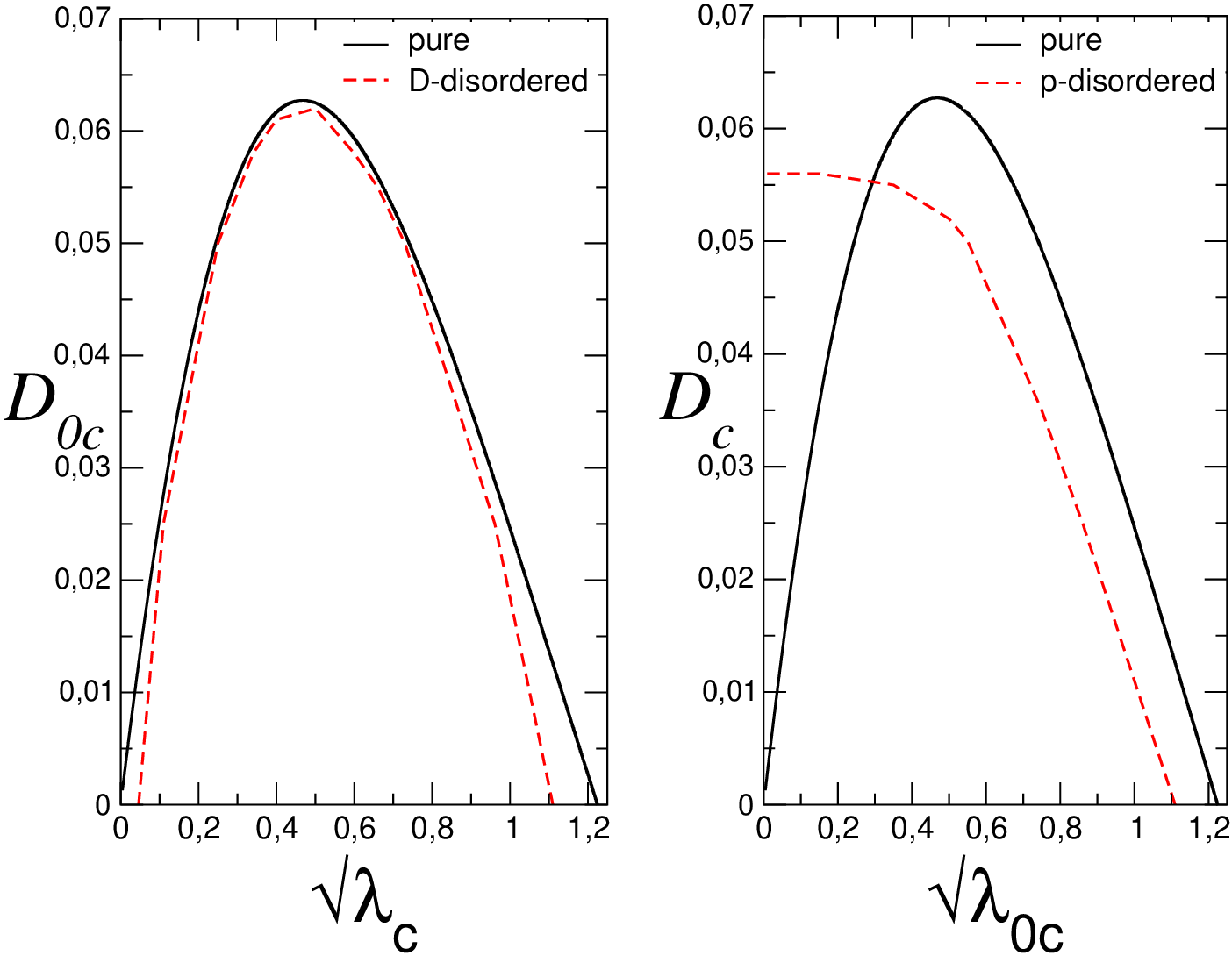}
\caption{({\bf Color online}): Two site MFT phase diagram for
the pure (full line) and disordered (dashed lines) TAM.
Panels $(a)$ and $(b)$ show results for the $D-$ and $p-$
disordered  for $\sigma=0.05$ and $0.5$, respectively.}
\label{pen}
\end{figure}

Aimed at to comparing with numerical simulations,
we accomplish mean field approximations (MFT) for temporal
disorder TAM.
Since the diffusion conserves the particle number,  
it is required to take into account at least correlations of two sites
to have included its effect. 
The time evolution of quantities are identical to those obtained by Dickman \cite{dic89}, but
instead the control parameter becomes time dependent. We focus on the uniform distribution, 
in which for  the $p-$ and $D-$disordered TAM  it follows that 
$\lambda(t)=\lambda_0+(2\xi-1)\sigma$ and $D(t)=D_0+(2\xi-1)\sigma$, respectively.
In Fig. \ref{pen},  the phase diagrams
for are shown for  $\Delta t=1$ and
 $\sigma=0.05$ and $\sigma=0.5$, for the $D-$ and $p-$disordered TAM,
respectively. For comparison results for 
the pure two-site MFT model (continuous lines) are included. 

Note that in all cases, the two-site MFT reproduces partially the numerical
findings. Whenever the reentrance is suppressed for the $p-$disordered TAM, it is preserved
for $D-$case. However there are some differences between MFT and
numerical results. In both cases, the
critical lines are encompassed  by the pure ones. This feature is verified 
numerically only for the $p-$case and larger diffusion rates.  We believe that these
discrepancies should be eliminated by taking into 
account  higher level of correlations in the time equations 
than those considered here.

\section{Comparison with the contact process}
\begin{figure}
\includegraphics[scale=0.35]{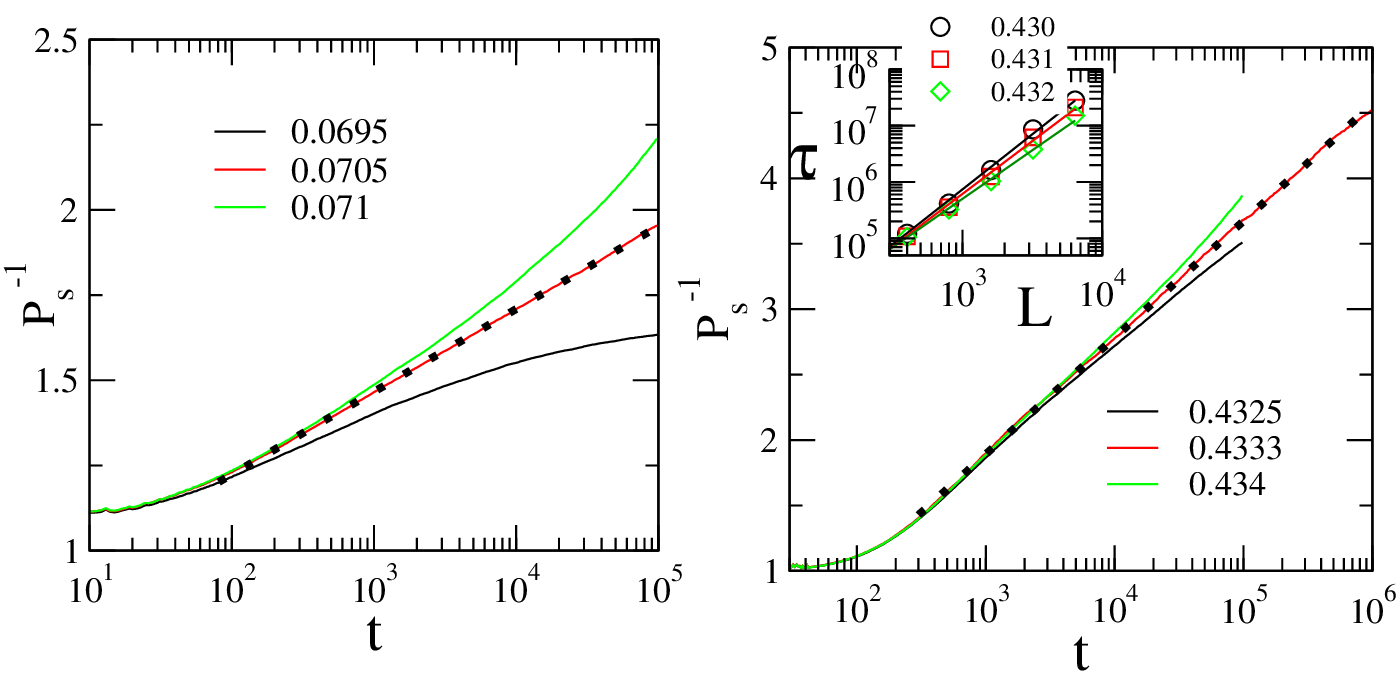}
\caption{({\bf Color online}): Results for the $p-$disordered
CP for a bimodal distribution.
Panels $(a)$ and $(b)$ show results for $D=0$  and $D=0.95$,
 respectively. In the inset, the log-log plot of $\tau$ versus $L$. }
\label{figcp}
\end{figure}

 In order to achieve a complementary information about
the competition  between diffusion and temporal disorder, in the last section 
 we present a brief analysis for the usual contact process.
We focus on the $p-$disordered case with a  bimodal distribution, whose 
numerical simulations have been performed
in a similar way than for the TAM and from Refs. \cite{neto,neto2}.
Fig. \ref{figcp} shows
the main results for $D=0$ and $0.95$.
In all cases, we take $\alpha_h=10 \alpha$, $\Delta t=6$ and ${\tilde p}=0.2$.

In both cases, results follow very well the strong noise RG \cite{neto,neto2}, even
for sufficient large diffusion values (exemplified for $D=0.95)$. 
Results for $D=0.99$ (not shown) still  exhibit a similar behavior, but
for only larger times. Also,  algebraic dependence
on the life time $\tau$  for finite systems (inset)
in the active phase ($\alpha < \alpha_c \sim 0.433$) with
non-universal exponents 
are  presented for $D=0.95$.  Thus,  contrasting to the TAM, 
above results suggest that the inclusion of strong  
diffusion in the CP  brings less pronounced changes for the TAM.  Therefore,
the influence of diffusion in  
both models seems leading to complementary features. 
However, we remark that more studies are still required, including 
the influence of $\Delta t$ or other kinds of disorder.

\section{Discussion and conclusions}
In this paper we give a further step for
establishing the effects of temporal disorder in one-dimensional
absorbing phase transitions. Aimed at addressing universality
hypothesis, the competition between disorder with diffusion 
and inspecting different kinds of disorder, we investigated a variant of the CP process, 
holding a reentrant phase transition and active phase for sufficient low
activation rates. 
Our conclusions can be summarized as follows: 
(i) For the uniform distribution, the disorder introduced in the creation-parameter or in the
diffusion induces slightly different phase diagrams. While the reentrant
shape is suppressed in the former, it is enlarged under the latter approach. On the other hand,
for bimodal disorder distribution, the reentrant portrait is maintained.
(ii) The role of diffusion is different in both cases, leading to 
distinct critical behaviors. Our findings  suggest
that whenever the $p-$disordered case  approaches to the DP
universality class as the diffusion increases (for both distributions),  for the
$D-$disordered case, the  DP
behavior is restored only in the reentrant 
part (at least for larger $\sigma$'s). 
 Since they contradict the Kinzel criterion about the relevance of
temporal disorder,   a crossover to
an universality class other than DP on much larger time scales 
is not excluded in such cases.
Also, 
in order to draw a comparison
with numerical findings, analysis
in the level of two site MFT
has been performed. In all considered cases,
the phase diagrams are in part similar with those obtained numerically.
(iii) For the $p-$disordered case, as the temporal disorder 
increases (by raising $\sigma$ or/and $\Delta t$), 
our results corroborate the strong-disorder theory by Vojta et al. 
\cite{neto,neto2}. However, results for the TAM
indicate that the effect of diffusion is more "drastic" than
for the CP.

Although numerical results indicate a critical behavior consistent to
 the usual algebraic behavior for weaker disorder,
 we believe that further studies are still required,
in order to address the critical behavior under  firmer basis. 
 In other words,
a crossover to the regime from Eq. (\ref{eq3}) for larger times  is possible
in such cases.

Therefore  this work  provides evidences that phase diagram and the critical behavior 
of {\emph low dimensional} systems with absorbing states can be different depending on 
the way the temporal disorder is introduced. This result can be interesting in the analysis of
real systems subjected to environmental noise.

\section*{ACKNOWLEDGMENT}
We acknowledge J. A. Hoyos for a critical reading of
this manuscript and useful comments.  
The financial supports from CNPq and FAPESP, under grants 15/04451-2
and 307620/2015-8, are also acknowledged.
\bibliographystyle{apsrev}

\end{document}